\documentclass{aa} 
\usepackage{graphics} %
\begin{document}

   \thesaurus{07         
	      (02.16.2;  
	       02.13.3;  
	       08.03.2;  
	       08.09.2 CU Vir;  
	       08.13.1;  
	       13.18.5)  
		}
   \title{Coherent radio emission from the magnetic chemically 
peculiar star CU Virginis}
  
   \author{C. Trigilio\inst{1,2}
	\and P. Leto\inst{1} 
	\and F. Leone\inst{2}
	\and G. Umana\inst{1,2}
	\and C. Buemi\inst{1}
	  }
 
   \institute{ 
Istituto di Radioastronomia del C.N.R., P.O. Box 141,
	      I-96017 Noto (SR), Italy
\and Osservatorio Astrofisico di Catania, Citt\`a Universitaria,
	      I-95125 Catania, Italy
	      }
   \offprints{C. Trigilio}
   \mail{trigilio@ira.noto.cnr.it} 

   \date{Received, accepted}

   \titlerunning{Coherent radio emission from CU Virginis} 
   \authorrunning{C. Trigilio et al.}
   \maketitle

   \begin{abstract}
Radio observations of the magnetic chemically peculiar star \object{CU~Vir}, 
carried out with the VLA in three different days, show that the radio emission 
at 20~cm is characterized by a strong enhancement at 
particular rotational phases. This radio emission is found to be right hand 
polarized with a degree of polarization close to 100~\%. 
As common for this class of stars, the magnetic axis of \object{CU~Vir} is 
oblique with respect to the rotational axis. By comparing the 20~cm radio
light curve with the effective magnetic field available from the literature, 
a coincidence of the main peaks of the radio emission with the magnetic nulls 
has been found. This happens when the magnetic axis lies in the plane of the 
sky.~\\
We suggest that the high degree of polarization, together with the high 
directivity of the radiation, can be explained in terms of coherent radio 
emission.
The data have been interpreted on the basis of the Electron Cyclotron
Maser Emission from electrons accelerated in current sheets out of the
Alfv\'en radius toward the stellar surface and eventually reflected 
outward by magnetic mirroring.
      \keywords{ 
	Stars: chemically peculiar --
	Stars: individual: CU Vir --
	Polarization --
	Stars: magnetic field --
	Radio continuum: stars --
	Masers
	}
   \end{abstract}
%
\begin{figure*} 
\resizebox{\hsize}{!}{\includegraphics{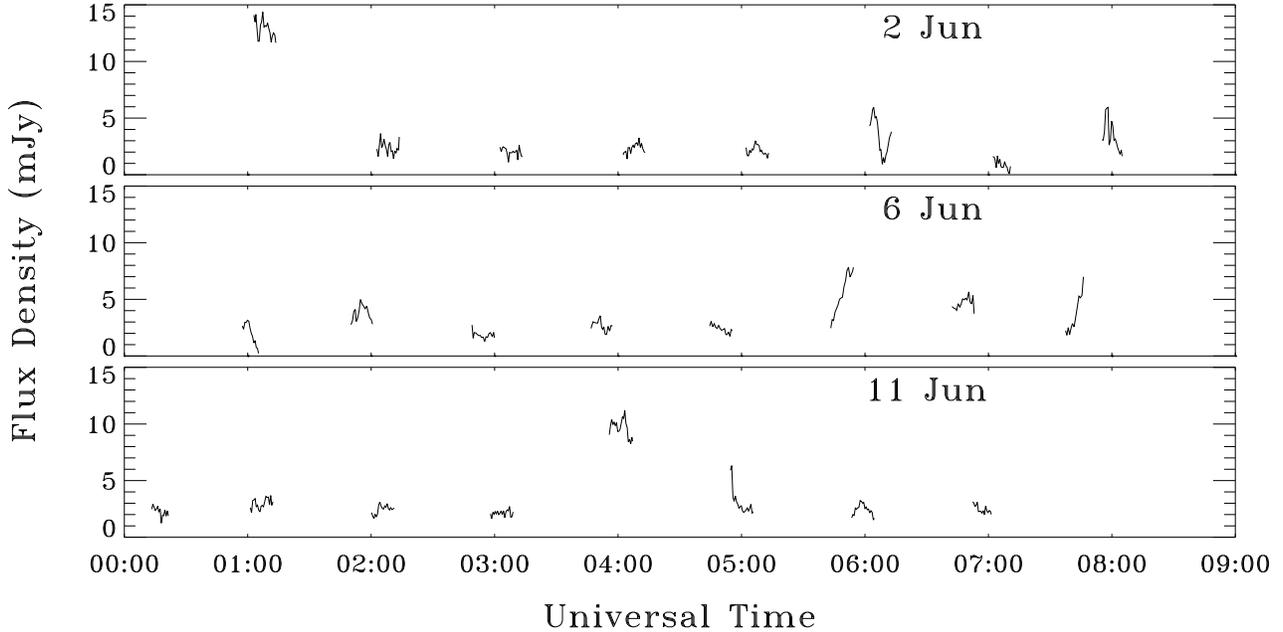}} 
      \caption{Flux density (Stokes I) at 1.4~GHz as a function of time in
      the three days of observation. Strong enhancements of the radio
      emission are evident.
	}
  \label{times}
\end{figure*}
\section{Introduction}
Magnetic chemically peculiar (MCP) stars can produce radio emission at  
centimetric wavelengths. This radio emission is consistent
with gyrosynchrotron emission from continuous ejected, mildly relativistic
nonthermal particles trapped in the magnetosphere (Drake et al. \cite{drake}).

MCP stars are characterized by mainly dipolar magnetic fields, 
whose strength steeply decreases with the stellar distance, and thus each 
radio frequency is expected to be emitted in a well localized stellar shell of 
the circumstellar region. Since the dipole axis is tilted with respect to
the rotation axis, Leone (\cite{leo91}) suggested that the observed radio 
emission from MCP stars should be periodically variable. Combining their own 
observations with Drake et al. (\cite{drake}) and Phillips \& Lestrade 
(\cite{phil}) data, Leone \& Umana (\cite{leo93}) have shown that the 6 cm 
density fluxes of \object{HD\,37017} and \object{HD\,37479} vary with the 
stellar rotation period. 
The coincidence of radio maxima with the extrema of the effective 
magnetic field lead Leone \& Umana to suggest that radio emitting regions are 
located above the magnetic poles. The variability of the 6 cm emission with 
the rotational period and has been also observed for the MCP star
\object{HD\,133880} by Lim et al. (\cite{lim}).

We have monitored along the rotational period \object{CU~Vir} (= HD\,124224) 
with the Very Large Array (VLA) at four frequencies. 
This star is particularly suitable for studying the radio emission from MCP 
stars because of its small distance (80 pc), very short rotational period 
(0.52 days) and reversing magnetic field. 
Here we present the behavior of the 
1.4\,GHz radiation whose properties are compatible with cyclotron maser 
emission.

\section{Observations and data reduction}
We have monitored at 1.4, 5, 8.4 and 15~GHz the B9p \ion{Si}{} star 
\object{CU~Vir} (= HD~124224) over three separate days using the VLA
\footnote{The Very Large Array is a facility of the National Radio Astronomy 
Observatory which is operated by Associated Universities, Inc. under 
cooperative agreement with the National Science Foundation}.
The observations have been carried out on June 1, 6 and 11, 1998 from 23 to 07 UT with all the available telescopes at the beginning of the A to B 
reconfiguration. For each frequency, we used the standard observing mode with 
two independent 50~MHz bands in Right and Left Circular Polarizations (RCP and 
LCP), adopting a 10 sec integration time.
At 1.4~GHz the two bands are separated by 80~MHz, being centered at 1385 and 
1465~MHz.

A typical observing cycle consisted of 10-min on source preceded and 
followed by 2-min on the phase calibrator \object{1354$-$021}, which is only
$6\degr$ far from \object{CU~Vir}. 
The four frequencies have been observed alternately with all the telescopes,
so that at each frequency \object{CU~Vir} has been observed for about 
$1^\mathrm{h}20^\mathrm{m}$ over a total time of $8^\mathrm{h}$.
The sequence of the frequencies during the three observing runs have been
organized to get the best possible sampling of the rotational phases,
avoiding redundancies.
To get a reliable flux scale, the amplitude calibrator \object{3C286} was 
observed at the beginning and the end of each run ($40\degr$ and $35\degr$ of 
elevation respectively).

Data were calibrated and mapped by using the standard procedures of the 
Astronomical Image Processing System (A.I.P.S.).
\object{CU~Vir} was found unresolved on VLA baselines at all the
observed frequencies.
Its position was determined by using the task JMFIT,
and found to be coincident with the position given in the Hipparcos 
Catalogue (Perryman et al. \cite{hip}):
$\alpha(2000) = 14^\mathrm{h}12^\mathrm{m}15\fs 8$ and
$\delta(2000) = +2\degr 24\arcmin 34\farcs 0$.
The temporal variation of the Stokes I and V parameters 
was determined with the task DFTPL.
This task performs the direct Fourier transform of the visibilities
as a function of time for an arbitrary position in the map. 
The results of this task can be affected by the sidelobes of any other strong 
source in the field. However, since the only source close to our target 
is very weak, having a flux density of only 3~mJy at 1.4~GHz, 
the possible confusion is negligible.
Heliocentric correction was then applied to the times of observation. 
Fig.~\ref{times} shows the radio light curve at 1.4~GHz during the three days
of observation.
\begin{figure*} 
\resizebox{\hsize}{!}{\includegraphics{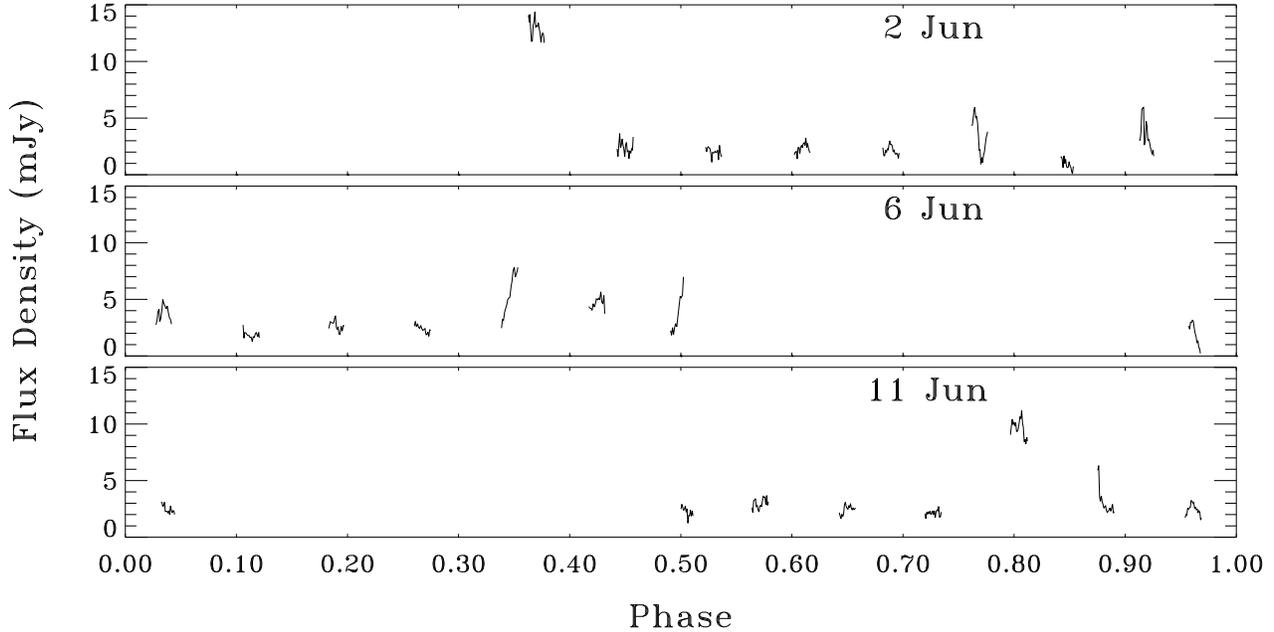}} 
      \caption{Flux density (Stokes I) at 1.4~GHz as a function of rotational
      phase, computed with the Pyper et al. (1998) ephemeris, in the three
      days of observation.
	}
  \label{phases}
\end{figure*}
%

\begin{table}
   \caption[]{Stellar parameters}
    \label{param}
      \[
      \begin{array}{p{0.5\linewidth}ccc}
      \hline
           \noalign{\smallskip}
spectral and 
peculiarity type      & \mathrm{SP} & \mathrm{B9 Si}  &                   \\
magnitude             & \mathrm{V}  & 5.01            &                   \\
distance              & d         & 80 \pm 6          & \mathrm{pc}       \\
radius                & R_\ast    & 2.2\pm 0.2        & R_{\sun}          \\
rotational period     & P         & 0.5207            & \mathrm{days}     \\
inclination           & i         & 43\degr\pm 7\degr &                   \\
obliquity             & \beta     & 74\degr\pm 3\degr &                   \\
polar magnetic field  & B_\mathrm{P} & 3000\pm 200    & \mathrm{gauss}    \\
          \noalign{\smallskip}
      \hline
     \end{array}
    \]
\end{table}

\begin{figure} 
\resizebox{\hsize}{!}{\includegraphics{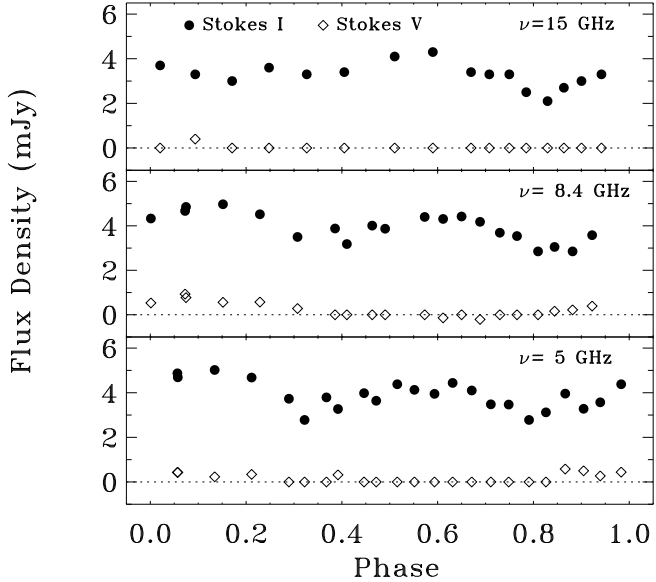}} 
      \caption{Radio emission at 15, 8.4 and 5~GHz during the three days of
      observation as a function of the rotational phase. No strong
      enhancement of the emission has been observed at any of those
      frequencies, rather a modulation is evident. Moderate circular
      polarization (Stokes V) has been detected at 8.4 and 5~GHz, with a
      percentage not greater than 10\%, in the range of phases 0--0.3
      and 0.8--1, slowing variating with the phases.
	}
  \label{uxc}
\end{figure}

\section{Stellar parameters}
Spectral, photometric and magnetic variations of MCP stars are characterized
by a single period. Stibbs (\cite{stibbs}) proposed that these stars present a 
dipolar magnetic field, whose symmetry axis is tilted with respect to the 
rotational axis, and a non homogeneous distribution of chemical elements on the 
stellar surface. Thus the variability period is the rotational one.

The rotational period of \object{CU~Vir} was firstly determined by Deutsch 
(\cite{deutsch}) equal to 0.52067 days. Several authors have later re-determined
this period with values ranging from 0.52067 to 0.5207 days, 
see Catalano \& Renson (\cite{cat}) for a complete list of references.
Recently, Pyper et al. (\cite{pyper}) analyzed all the photometric data
covering 40 years. They suggested that the rotational period of \object{CU~Vir}
increased abruptly between 1983 and 1987
of a factor $\Delta P/P\approx 4.9\times 10^{-5}$. This unexpected jump
of the rotational period was discussed by St\c{e}pie\'n (\cite{stepien}). 
He suggested that the spin down occurs in the outher stellar 
envelope alone, due to a change of the moment of inertia of the envelope
itself because of the internal magnetic field strength.
The new ephemeris given by Pyper et al. (\cite{pyper}), referred to light 
minimum, are:\\

\noindent 
$HJD=2435178.6417+\left\{ 
	\begin{array}{lr}
	0^d.52067780 E & \mbox{JD $<$ 2\,446\,000} \\
	0^d.52070308 E & \mbox{JD $>$ 2\,446\,000} 
	\end{array}
\right.
$\\

Using the projected rotational velocity 
$v_\mathrm{e}\sin i = 146\pm 2\, \mathrm{km\,s^{-1}}$ 
(Hatzes \cite{hatzes}) and the stellar radius 
$R_{\ast}=2.2 R_{\sun}$ (North \cite{north}), in the relation
$$ \sin i = P_\mathrm{days} 
          \frac{v_\mathrm{e}\sin i}{50.6}
          (R_\ast/R_{\sun})^{-1}, $$
valid for a rigid rotator,
we measure an inclination of the rotational axis with respect to the line 
of sight $i = 43\degr \pm 7\degr$. 

Borra \& Landstreet (\cite{borra}) found that the effective 
magnetic field ($B_\mathrm{eff}$), measured from 1976 to 1978, is variable
with the period established by Winzer (\cite{winz}) from photometry.
$B_\mathrm{eff}$ is the average of the component of the local magnetic
field along the line of sight over the whole stellar disk.
From the value of the inclination $i$ and Borra \& Landstreet (\cite{borra}) 
measurements of $B_\mathrm{eff}$, we obtain an obliquity of the dipole
$\beta = 74\degr\pm 3\degr$ and a polar magnetic field equal
to $B_\mathrm{p} = 3000\pm 200$ gauss. 
Pyper et al. (\cite{pyper}) give new values of $B_\mathrm{eff}$; however,
we prefer to infer the magnetic geometry from Borra \& Landstreet
measurements which are based on hydrogen lines. In fact, Pyper and co-workers
measurements are based on the 634.7 and 637.1 nm lines of silicon 
and this element is  
not homogeneously distributed on the surface of \object{CU~Vir}
(Kusching et al. \cite{kusching}). Even if we cannot exclude a more complex
topology of the magnetic field, like a dipole+quadrupole or a decentered 
dipole, as proposed by Hatzes (\cite{hatzes}), the assumption of a simple 
dipole at an height above the photosphere where the radio emission is 
generated seems reasonable, as the strength of the quadrupole field decreases 
one order of magnitude faster than the dipole field. 

\noindent A summary of the adopted and derived parameters for \object{CU~Vir} 
is reported in Table~\ref{param}.

\begin{figure} 
\resizebox{\hsize}{!}{\includegraphics{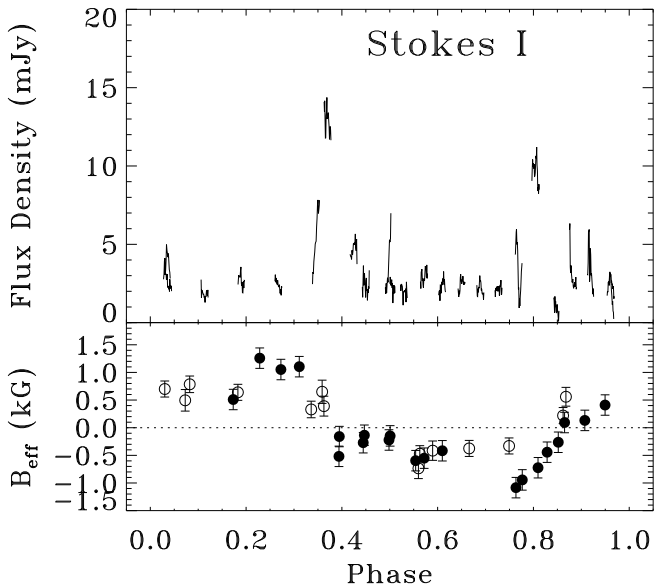}} 
      \caption{
	Upper panel: 
		1.4\,GHz, emission from \object{CU~Vir};
	Lower panel: 
		Longitudinal magnetic field measurements 
		from Borra \& Landstreet (1980) and from Pyper et al. (1998)
		(open and filled circles respectively) as a function of the 
		rotational phase.
                1.4\,GHz radiation maxima correspond approximately with the
                null magnetic field.
	}
  \label{borra}
\end{figure}
 
\section{The 1.4\,GHz emission}
The observed 1.4\,GHz emission is shown in Fig.~\ref{phases}  
as a function of the phase during the three days of observation separately.
In Fig.~\ref{uxc} a summary of the behavior of other three frequencies
as a function of the phase for the whole observing run is reported.
The smooth variations of the radio flux and of the circular polarization
at the higher frequencies will be discussed elsewhere, together with a
numerical model for the continuum radio emission from MCP stars.

\subsection{Correlation with the magnetic field}
Since Leone \& Umana (\cite{leo93}) reported a correlation between 
radio emission and effective magnetic field in \object{HD\,37017} and 
\object{HD\,37479}, we look first at a similar correlation for our data.
Fig.~\ref{borra}, upper panel, shows the 1.4~GHz emission as a function of
the rotational phase. 
This emission is characterized by two or three components: a basal flux 
of $2-3$~mJy and very large increments of the flux around phases
0.35-0.45 and 0.75-0.85, where it goes up to 15 mJy, and secondary peaks 
of up to 7~mJy, further discussion of which is deferred to Section 5.2.2.
Only in two moments, at phase 0.85 and 0.97, the intensity drops below 1~mJy.
Phasing the magnetic data from Borra \& Landstreet (\cite{borra}) and Pyper 
et al. (\cite{pyper}) (Fig.~\ref{borra}, lower panel), we found that 
those peaks of the radio emission coincide approximately with the the null
magnetic field, i.e. when the axis of the dipole is almost perpendicular
to the line of sight. The peak visible at phase 0.35-0.45 is defined by
the observations on June 2 (maximum) and 6 (rising phase), and the 0.75-0.85
one on June 2 (rising phase) and 11 (maximum) (see also Fig.~\ref{phases}).
According to the accuracy of the period given by Pyper et al. (\cite{pyper}),
our data are phased with respect to the magnetic data better than 
$3\times 10^{-3}$.

It is worthy to note that the flux density of the peaks is about five times 
larger than those previously reported in the literature at this frequency for 
MCP stars. Previous observations of \object{CU~Vir} reported by Leone et al. 
(\cite{leo96}) show that the 1.4~GHz flux is 2.6~mJy, in agreement with the
"out of peaks" emission here reported.

The high flux increment, that occurs at the particular orientation of the
magnetosphere, indicates an emission mechanism that is not explainable
with the emission models up to now  proposed. In the following, we will 
analyze further characteristics of this emission. 

\begin{figure} 
\resizebox{\hsize}{!}{\includegraphics{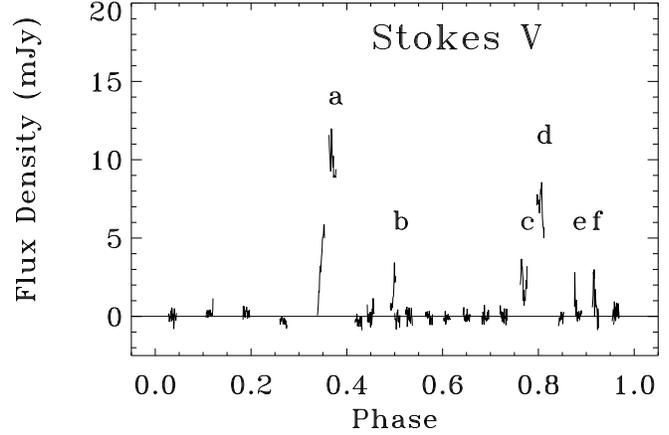}}
      \caption{Stokes parameter V versus rotational phase in the 3 days of 
	observation. No polarization is seen outside the peaks denoted with the 
	letters from a to f. The higher fluxes ({\it a} and {\it d}) are 
	visible when the magnetic axis is perpendicular to the line of sight.}
  \label{r-l}
\end{figure}
\subsection{Polarization}
A further exceptional aspect of the 1.4~GHz is its high degree of circular
polarization. 
Fig.~\ref{r-l}, where the Stokes V parameter (V = 1/2(RCP$-$LCP)) is
plotted versus the rotational phase, shows that the
increment of the flux occurs only in the right-hand circular polarization.
Six peaks, denoted with the letters from {\it a} to {\it f}~ in Fig.~\ref{r-l}, 
are visible, the largest and broad emission being detected in peaks {\it a}
and {\it d}, that occur in coincidence with the null magnetic field.
Outside the peaks, the Stokes parameter V is statistically null. 
The percentage of polarization $\pi_\mathrm{c}$ goes up to 80~\% during the 
main peaks {\it a} and {\it d}. If we subtract the continuum radio emission,
that contributes for about 3~mJy, we get $\pi_\mathrm{c} \approx 100$~\%.

\begin{figure} 
\resizebox{\hsize}{!}{\includegraphics{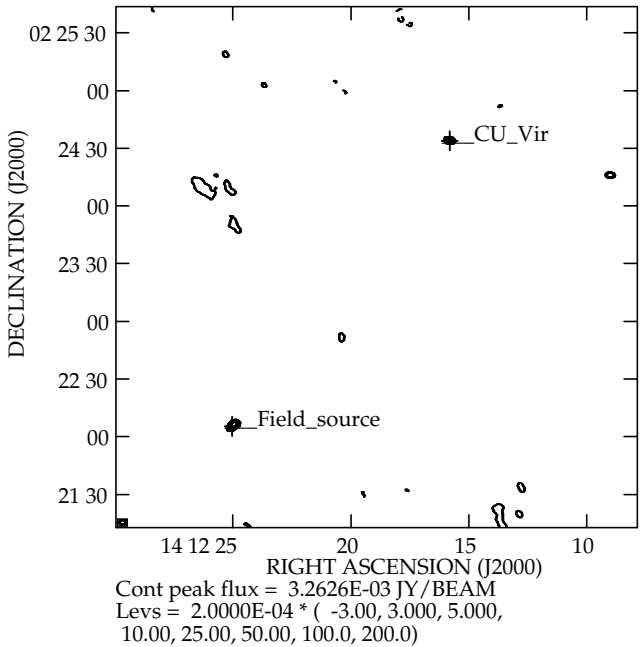}}
      \caption{The 1.4\,GHz radio map of \object{CU~Vir} showing the field 
	source used to check possible instrumental effects.}
  \label{mappa}
\end{figure}

\begin{figure}
\resizebox{\hsize}{!}{\includegraphics{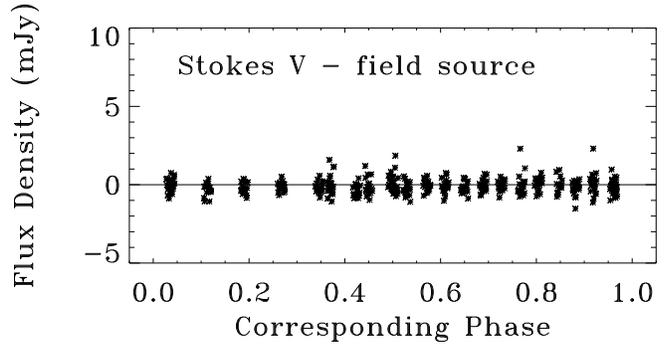}}
      \caption{Stokes V for the field radio source as function of 
	the corresponding rotational phase of \object{CU~Vir}. The Stokes I
	is 3~mJy. The absence of variability in the polarization confirms that 
	no instrumental problem occurred during the observations.}
  \label{confronto}
\end{figure}

In order to verify if the behavior of the right-hand circular polarization
was not due to an instrumental problem, we looked for other sources in the 
field of our star to be monitored for the polarization at the same times of 
our observations.
Leone et al. (\cite{leo96}) found in the 6\,cm VLA frames centered at the
\object{CU~Vir} position a radio source with coordinates:
$\alpha(2000) = 14^\mathrm{h}12^\mathrm{m}24\fs 96$ and
$\delta(2000) = +2\degr 22\arcmin 04\farcs 7$, that they used to check possible
instrumental effects. This field source is visible also in the 1.4~GHz map 
showed in Fig.~\ref{mappa} and has a flux density (Stokes I) of about 3~mJy, 
comparable with \object{CU~Vir} outside the peaks. 
By comparing the polarizations (Stokes V) of the field source and of 
\object{CU~Vir}, we can rule out any instrumental effects 
(Fig.~\ref{confronto}). 

\subsection{Directivity of the radio emission}
Since the polarized peaks have been observed in the three different days 
of observation, we can conclude that this radiation is stable at least in 
a period of weeks.
The observed variations can be due to the different inclination that the
oblique dipole forms with the line of sight as the star rotates. 
In a particular geometric configuration, highly beamed radiation is emitted 
toward the Earth, producing the observed peaks.
As already pointed out, the main peaks of the right-hand circular polarization
corresponds almost to the zero of the average magnetic field over the surface 
of the star. This means that the maxima of this emission occur when the 
magnetic axis is almost perpendicular to the line of sight. 

\begin{figure*} 
\resizebox{\hsize}{!}{\includegraphics{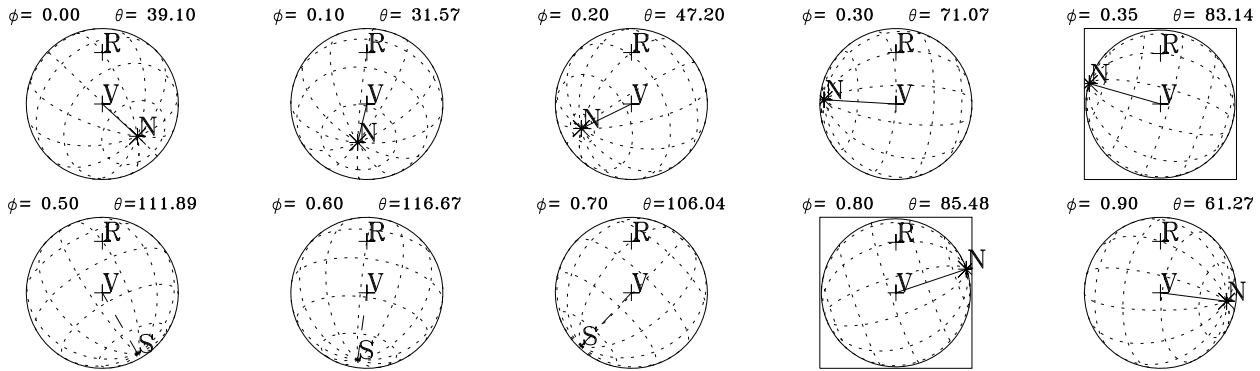}} 
      \caption{Changes of the geometry of CU Vir as a function of the
      rotational phase. The point V at the center of the star is the
      intersection of the stellar surface with the line of view; the point P
      close to the top is the rotational pole; the points denoted with N
      and S represent the North and South magnetic pole; the arc V-N 
      subtends the angle $\theta_\mathrm{M}$. 
      $\phi$ and $\theta_\mathrm{M}$ are indicated
      in each panel. The main peaks are visible when the star is oriented
      as in the two framed panels at $\phi\approx 0.35$ (peak $a$) and
      at $\phi\approx 0.80$ (peak $d$).}
  \label{cartoon}
\end{figure*}

Borra \& Landstreet (\cite{borra}) found that the magnetic curve is delayed
with respect to the light curve. 
Assuming that the two main peaks occur in two symmetric orientations of the
magnetosphere, we can compute this shift with good accuracy: 
from our data, peaks $a$ and $d$ show their maxima at phases
$\phi_\mathrm{a}=0.37$ and $\phi_\mathrm{d}=0.80$ respectively, and therefore
$$
\phi_0 = (\phi_\mathrm{a}+\phi_\mathrm{d})/2-0.5 = 0.08.
$$ 

Once the geometry of the magnetosphere of the star has been defined by the
values of the inclination $i$ and the obliquity $\beta$ (see Table~\ref{param}),
we can find the angle $\theta_\mathrm{M}$ that the line of sight forms with 
respect to the axis of the dipole, defined by:
$$
\cos\,\theta_\mathrm{M}=
\sin\,\beta\,\sin\,i\,\cos(\phi-\phi_0)+\cos\,\beta\,\cos\,i.
$$
The error of $\theta_\mathrm{M}$ estimated from this relation and from
the uncertainty in the angles $i$ and $\beta$ is about $4\degr$ when 
$\theta_\mathrm{M}\approx 90\degr$ and $8\degr$ when 
$\theta_\mathrm{M}\approx 30\degr$ and $120\degr$. 
Fig.~\ref{cartoon} shows the change of the orientation of the star, and so 
of the angle $\theta_\mathrm{M}$, with the rotation.

\begin{figure}
\resizebox{\hsize}{!}{\includegraphics{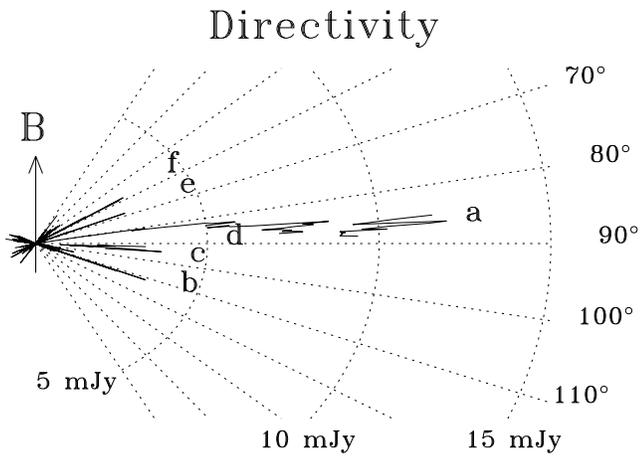}}
  \caption{Polar representation of the emission, showing the very high 
	directivity of the coherent components. The arrow B represents the
	axis of the dipole; the emission is plotted as a function of the
	angle $\theta_\mathrm{M}$ formed by the line of sight and the
	axis B.}
   \label{beam}
\end{figure}

Fig.~\ref{beam} shows the behavior of the polarized component of
the emission as a function of the angle $\theta_\mathrm{M}$ in a polar
view. The main peaks are emitted at an angle of about $85\degr$ with 
respect to the axis of the dipole (indicated as the arrow B), and have a high 
degree of directivity, with an half power beam width of about $5\degr$.
For the less intense peaks $b, c, e$ and $f$, the beams have a narrower width 
of about $1\degr$.
Fig.~\ref{main} shows the details of the main peaks as a function of the angle 
$\theta_\mathrm{M}$.
It is important to note that the magnetic longitude of the line of sight
during the two main peaks are different, since in these configurations
the star shows opposite hemispheres (see Fig.~\ref{cartoon}). This means that
the observed phenomenon is not related to an active longitude, depending
only on $\theta_\mathrm{M}$. It seems to be similar to the pulses observed
in pulsars, the magnetic fields of which are, like MCP stars, typically
characterized by oblique dipole geometries.
\section{Discussion}
To explain the radio continuum and the X-ray emissions from young magnetic 
B stars and Bp-Ap stars, Andr\'e et al. (\cite{andre}) proposed a model  
where the stellar wind plasma flows out near the magnetic poles along the 
field lines. 
Far from the star, such a wind draws the field lines near the equator into 
current sheets which should be location of particle acceleration. 
Mildly relativistic electrons return to regions near the star
by traveling along the magnetic field lines, emitting gyrosynchrotron 
radiation. This model has been applied by Linsky et al. (\cite{linsky}) to
explain the flat spectra of the MCP stars.

The characteristics of the polarized component of the radio emission at 1.4~GHz
from \object{CU~Vir}, that show high degrees of circular polarization and 
high directivity, can be explained in terms of coherent emission. 
The two major mechanisms that have been suggested to explain coherent radio
emission are plasma radiation due to Langmuir waves and cyclotron maser.
Plasma radiation has been invoked in several cases, such as solar microbursts
at 1.4~GHz (e.g. Bastian \cite{bastian}), solar millisecond spikes (e.g.
Wentzel \cite{wentzel}) and radio bursts from the flare star \object{AD~Leo}
(Abada-Simon et al. \cite{abada}).
The theory of Electron Cyclotron Maser Emission (ECME) (e.g. 
Wu \& Lee \cite{wu} or Melrose \& Dulk \cite{meldulk}) seems to be the 
favourable emission mechanism to explain coherent radiation from the 
magnetosphere of Jupiter, solar spike bursts and flare stars, like M dwarfs 
and RS~CVn binary systems.

\subsection{Plasma Radiation}
Plasma radiation is a two-stage process where longitudinal waves in the 
plasma (Langmuir waves) are first generated and later their energy is converted 
into radiation (e.g. Dulk \cite{dulk}). The frequency of the radiation is the 
plasma frequency ($\nu_\mathrm{P}\approx 9\,000 \sqrt{N_{\rm e}}$~Hz, 
with $N_{\rm e}$ the plasma density number in $\mathrm{cm}^{-3}$)
or its second harmonic. 
To be observed at 1.4~GHz, $N_{\rm e}\approx 2.4\times 10^{10}$ or 
$6\times 10^9 \mathrm{cm}^{-3}$ respectively for $s=1,2$.
Plasma radiation can occur when the magnetic field is
relatively weak ($\nu_\mathrm{P} \geq \nu_\mathrm{B}$, where 
$\nu_\mathrm{B}=2.8\times 10^6B$, with $B$ in gauss). 
This gives an upper limit for the magnetic field $B$ in the region where the radiation is generated: $B < 500$ ($s=1$) or $B < 250$~gauss ($s=2$).
Assuming a dipolar topology of the magnetic field, $B=B_\ast(R_\ast/R)^3$, 
with $B_\ast=3\,000$~gauss, we get that the region where plasma emission occur
must be located at $R>1.8 R_\ast$ ($s=1$) or $R>2.3 R_\ast$ ($s=2$).

However, the theory of plasma radiation does not foresee any high directivity.
For this reason this mechanism is not suitable to explain our observations.

\subsection{Electron Cyclotron Maser Emission}

Following the theory of the ECME, electrons reflected by the magnetic 
mirrors can develop a pitch angle anisotropy (or loss cone anisotropy), 
becoming candidates for cyclotron maser emission if the local plasma frequency
is relatively small ($\nu_\mathrm{P} \ll \nu_\mathrm{B}$). 
The frequency of the maser 
emission is given by $\nu \geq s\nu_\mathrm{B}$, being $s$ the harmonic number 
and $\nu_\mathrm{B}$ the gyrofrequency. The faster growth rate is for the first 
few harmonic number ($s \leq 4$) in the extraordinary mode (x-mode).
The radiation generated where the magnetic field intensity is $B_0$ has a 
frequency $\nu_0 \approx s\nu_B$; when crossing a more external layer with 
$B_1=\frac{s}{s+1}B_0$, it can be suppressed by the gyromagnetic absorption
of the thermal plasma. 
Melrose \& Dulk (\cite{meldulk}) show that the $1^\mathrm{th}$ harmonic is 
generally suppressed, while the $2^\mathrm{th}$ can escape from layers 
absorbing at higher harmonic numbers. 
The ECME is confined in a hollow cone of half-angle $\theta$ with respect 
to the line of the magnetic field, with $\cos\,\theta =v/c$ and $v$ the speed
of the emitting electrons. 
The thickness of the hollow cone is $\Delta \theta \approx v/c$. 
If the maser is emitted in a region of constant magnetic field, 
the relative bandwidth is 
$\Delta \nu/\nu =\frac{\nu-s\nu_\mathrm{B}}
{s\nu_\mathrm{B}}\approx \cos^2\theta$.
However, if the emission comes from a layer where the magnetic field ranges from
$B_\mathrm{a}$ to $B_\mathrm{b}$, then the observed bandwidth will be much 
larger, ranging the radiation from $s\nu_\mathrm{a}$ to $s\nu_\mathrm{b}$. 
In this case, the angle $\theta$ and the thickness of the hollow cone of 
radiation $\Delta \theta$ do not change, depending only on $v/c$.
\begin{figure}
\resizebox{\hsize}{!}{\includegraphics{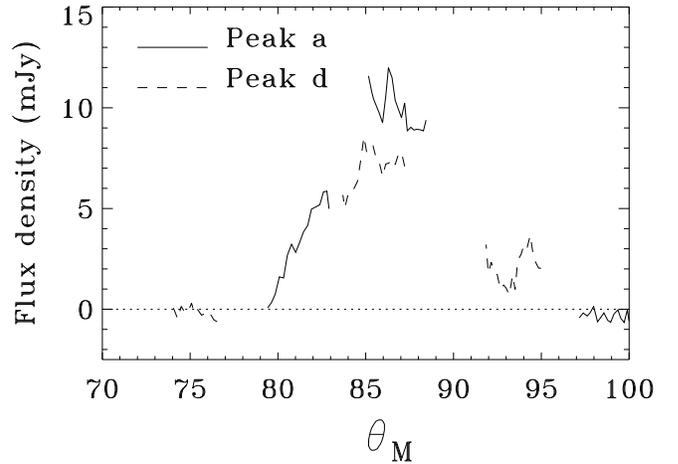}}
  \caption{The main peaks $a$ and $d$ as a function of the angle 
	$\theta_\mathrm{M}$.}
   \label{main}
\end{figure}
\begin{figure} 
\resizebox{\hsize}{!}{\includegraphics{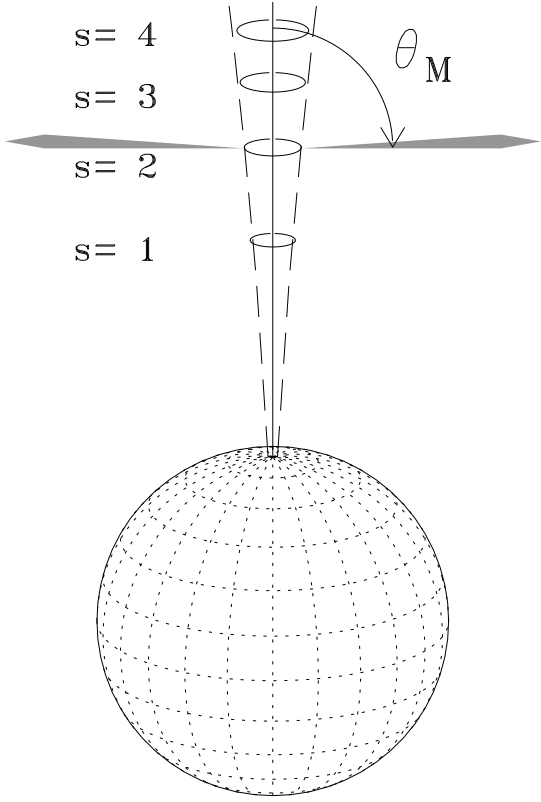}} 
      \caption{Proposed picture for the emission of the main peaks. 
      Accelerated in the current sheets and flowing back to the surface
      following the magnetic field lines (dashed lines from the pole)
      electrons are reflected by magnetic mirrors close to the star; 
      they develop
      a loss cone anisotropy and emit cyclotron maser when going outward;
      the location of the possible region of the maser emission at 1.4~GHz
      are the circular rings centered around the axis of the dipole
      (vertical straight line); each ring is marked with the corresponding
      harmonic number s. The shaded areas represent the escaping
      radiation if generated, for example, at harmonic s=2. The angle 
      $\theta_\mathrm{M}$ is also shown.
	}
  \label{emiss}
\end{figure}

\subsubsection{The main peaks}
We observe that the two main peaks $a$ and $d$ are beamed at an angle
$\theta_\mathrm{M} \approx 85\degr$ with respect to the axis of the dipole,
and have a full width half maximum (FWHM) of about $5\degr$ (Fig.~\ref{main}). 
This strongly suggests that we are in presence of ECME.

We propose the following scenario: the electrons accelerated in the current
sheets out of the Alfv\'en radius flow toward the photosphere close to the 
magnetic pole; they are eventually mirrored back by the increasing 
magnetic field, traveling along field lines almost parallel to the axis of
the dipole. After the reflection, they develop a loss cone anisotropy because 
of the interaction with the thermal plasma, leading to electron cyclotron maser
emission. 
In this hypothesis the angle $\theta_\mathrm{M}$ is just the angle $\theta$
of the ECME theory. Since $\Delta \theta \approx v/c$ and 
$\cos \theta \approx v/c$, we get $v/c \approx 0.09$. 

If the magnetic field of \object{CU~Vir} is a dipole, we should expect a 
symmetry in the ECME with the stellar phase. The absence of any beamed 
emission at $\theta_\mathrm{M}\approx 95\degr$ (i.e. at $85\degr$ from
the direction of the south pole) suggests an asymmetry of the magnetosphere.
The presence of a quadrupole component, that has been observed in other MCP
stars like \object{HD~32633} and \object{HD~175362} (Mathys \cite{mathys}), 
or of a decentered dipole, as recently proposed for \object{CU~Vir} 
(Hatzes \cite{hatzes}), could explain the observed asymmetry of the ECME.
In fact, the presence of a quadrupole, as for example shown by Michaud et al.
(\cite{michaud}) in their figure~1d, can inhibit the wind, and so the radio 
emission, from the magnetic south pole. The interpretation of an asymmetry
in the wind is not new. In fact Brown et al. (\cite{brown}), from the behavior
of the UV lines of the MCP star \object{HD~21699}, inferred that the wind
flows from "{\em only one} of the magnetic poles".
\begin{figure} 
\resizebox{\hsize}{!}{\includegraphics{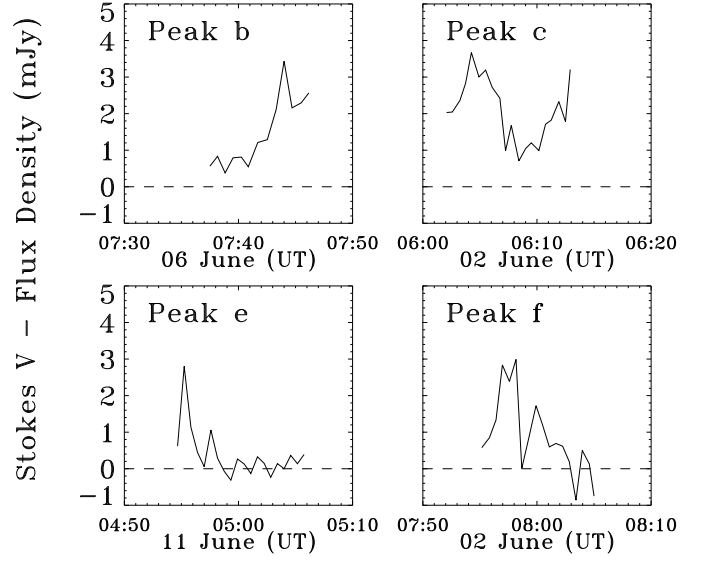}} 
      \caption{Secondary peaks. See text for details.
	}
  \label{secondary}
\end{figure}

\subsubsection{The secondary peaks}
The secondary peaks $b$, $c$, $e$ and $f$ are, like the main peaks $a$ and $d$,
also circularly polarized. 
The maximum flux density that they show is about 3~mJy in the Stokes V,
as it is possible to see inspecting Fig.~\ref{secondary}. Peak $b$ lasts for
2 minutes (half power to half power) and a further rise is possible. It has
been detect on June 6 at phase 0.5; at the same phase, on June 11,  
no flux enhancement has been observed, meaning that probably peak $b$
is a transient phenomenon. Peak $c$ lasts about 3--4 minutes and is followed
by the rise of the main peak $d$. Peak $e$ is very short in duration,
about one minute, and peak $f$ is about 4 minutes. 
While the main peaks $a$ and $d$ are detected in all the three days of 
observation, we cannot say if $b$, $c$, $e$ and $f$ are sporadic impulsive 
emissions or if they are stable as the main ones. Further observations are 
needed to clarify this point.

\subsubsection{The bandwidth}
Our observations have been performed at two bands of 50~MHz separated by
80~MHz. No difference of flux between the two bands has been found.
So, the bandwidth of the masing radiation is $\Delta \nu \gg 80\,\mathrm{MHz}$.
Since in the ECME theory 
$\Delta \nu/\nu \approx \cos^2\theta$, with $\theta=85\degr$ we
expect a bandwidth of about 10~MHz. This apparent incongruence can be 
explained if the region where the maser emission is generated covers a
wide range of magnetic field strength. The observed radiation is the
envelope of a continuous series of maser spots along the field lines,
the higher frequency being emitted in regions closer to the star, according to 
$\nu \approx s\nu_{\rm B}$. In a dipole $B=B_\ast(R_\ast/R)^3$; 
with $B_\ast=3\,000$~gauss, $\nu \approx 8.4\,s\,(R_\ast/R)^3$~GHz.
To be observed at 1.4~GHz, the maser spots are located at 
$R\approx1.8\,s^{1/3}$, with $s \leq 4$.
On the contrary, we did not observed any coherent emission at 5~GHz, as
we will discuss in a following paper. So, no condition
for the maser mechanism is expected close to the star. The maser mechanism
is efficient at a distance $R > 1.5\,s^{1/3}$. Probably the electrons
are thermalized close to the star, due to a higher density.

\subsubsection{Why only right hand circular polarization?}
The Stokes parameter V is always positive, that means the radiation is right 
hand polarized. 
The theory of the ECME foresees that the radiation is almost entirely polarized
in X-mode, as observed for the auroral kilometric radiation (AKR) and for the
Jupiter's decametric emission (DAM) 
(Wu \& Lee \cite{wu}, Melrose \cite{melrose}).
In fact, in the x-mode the sense of rotation of the electric vector
is the same as the helicity of the emitting particles.
Electrons moving in a magnetic field directed toward us are seen to rotate
in counter clockwise, as the right hand circular polarization.
In a perfectly symmetric configuration, the electrons mirrored outward
move in the same direction as the magnetic field lines in the north hemisphere,
in opposite direction in the south hemisphere, emitting respectively in right 
and left hand polarization. 

Our data show no LCP enhancement at any rotational phase. This suggests that
there is no condition for cyclotron maser emission at 20~cm in the magnetic 
south hemisphere. Again, this can be imputed to an asymmetry in the 
magnetosphere of \object{CU~Vir}.

\section{Conclusions}
The discovery of the coherent radio emission from \object{CU~Vir}, that
can be explained in terms of Electron Cyclotron Maser Emission, opens
new prospectives in the field of the magnetic chemically peculiar stars.

An important question that can be put is: 
{\em is the coherent radio emission at low frequency a common characteristic 
of the magnetic chemically peculiar stars?} Low frequency observations of a 
sample of MCP stars that present a reversal magnetic field have to be carried
out in order to answer this question. Such a survey will certainly put more
stringent parameters to draw the physical mechanisms that act in the 
magnetosphere of the MCP stars.

Further observations on \object{CU~Vir} are needed to confirm that the coherent
radio emission here reported is stable in time. In particular, a continuous 
coverage with the rotational phase to better define the angular distribution
of the coherent emission is crucial. Moreover, further observations covering
several rotational periods are also needed to study the nature of the secondary
peaks and to determine whether or not the phases at which they occur are truly
random.

The bandwidth of the radiation is found to be larger than 80~MHz. 
Observations at frequency lower than 1400~MHz are needed to better define
the spectrum.

\acknowledgements{We thank the referee, Dr. S. Drake, for comments and 
suggestions which enabled us to improve this paper.}

\end{document}